\documentclass{article}
\usepackage{amsmath}
\usepackage{amssymb}
\usepackage{mathrsfs}
\usepackage{graphicx}
\usepackage{float}
\usepackage{caption}
\usepackage{subcaption}

\usepackage{epsfig}
\usepackage{amsmath}
\usepackage{amsfonts}
\usepackage{graphicx}

\usepackage[margin=2cm]{geometry}

\begin{document}

\title{{\bf \Large  Kerr-Newman black holes with scalar hair}}

\vspace{0.5cm}

 \author{
 {\large Jorge F. M. Delgado}\footnote{jorgedelgado@ua.pt}, \
{\large Carlos A. R. Herdeiro}\footnote{herdeiro@ua.pt}, \
{\large Eugen Radu}\footnote{eugen.radu@ua.pt}  \ and
{\large Helgi R\'unarsson}\footnote{helgi.runarsson@ua.pt}
\\ 
\\
{\small Departamento de F\'\i sica da Universidade de Aveiro and} \\ 
{\small Center for Research and Development in Mathematics and Applications (CIDMA)} \\ 
{\small   Campus de Santiago, 3810-183 Aveiro, Portugal}
}


\date{August 2016}
\maketitle

\begin{abstract}
We construct electrically charged Kerr black holes (BHs) with scalar hair. 
Firstly, we take an uncharged scalar field, 
interacting with the electromagnetic field only indirectly,  via the background metric. 
The corresponding family of solutions, 
dubbed Kerr-Newman BHs with ungauged scalar hair, 
reduces to (a sub-family of) Kerr-Newman BHs in the limit of vanishing scalar hair 
and to uncharged rotating boson stars in the limit of vanishing horizon. 
It adds one extra parameter to the uncharged solutions: the total electric charge. 
This leading electromagnetic multipole moment is unaffected by the scalar hair 
and can be computed by using Gauss's law on any closed 2-surface surrounding  
(a spatial section of) the event horizon. 
By contrast, the first sub-leading electromagnetic multipole -- the magnetic dipole moment --, 
gets suppressed by the scalar hair, such that the gyromagnetic ratio is always smaller than the Kerr-Newman value ($g=2$).  
Secondly, we consider a gauged scalar field and obtain a family of Kerr-Newman BHs with gauged scalar hair. 
The  electrically charged scalar field now stores a part of the total electric charge, 
which can only be computed by applying Gauss' law at spatial infinity 
and introduces a new solitonic limit -- electrically charged rotating boson stars. 
In both cases, we analyse some physical properties of the solutions.
\end{abstract}



\section{Introduction}
The Kerr metric~\cite{Kerr:1963ud} is the fundamental BH solution in General Relativity, believed to describe an untold number of BHs in (or near) equilibrium in the Cosmos. It is straightforward to generalize this solution to include an electric (or magnetic) charge, yielding the Kerr-Newman (KN) solution~\cite{Newman:1965my}. The latter is, perhaps, of a more limited astrophysical interest, as electric charge is expected to be residual in astrophysical BHs, due to efficient discharge mechanisms (but see the discussion in~\cite{Cardoso:2016olt}). Theoretically, however, the KN solution introduces some qualitatively new features, with respect to its vacuum counterpart, including: a (Komar) energy and angular momentum component outside the horizon~\cite{Delgado:2016zxv}; a more general extremal limit with different (including supersymmetric, when embedded in supergravity~\cite{Gibbons:1982fy,Tod:1983pm,Herdeiro:2000ap}) properties, depending on the electric charge; and a dipole magnetic moment, induced by the electric charge and the rotation. The corresponding gyromagnetic ratio turns out to have precisely the (non-anomalous) electron value, $g=2$~\cite{Carter:1968rr}. Moreover, the KN solution provides an arena for the study of the fully non-linear interplay between electromagnetism and gravity, in the framework of Einstein's theory, with often challenging properties, as for instance its stability -- see, $e.g.$, the discussions in~\cite{Pani:2013ija,Pani:2013wsa,Zilhao:2014wqa}.

\bigskip

As recently discovered, the Kerr solution admits also families of generalizations with scalar hair~\cite{Herdeiro:2014goa,Herdeiro:2015gia,Kleihaus:2015iea,Herdeiro:2015tia,Chodosh:2015oma} (and also Proca hair~\cite{Herdeiro:2016tmi}). In its simplest guise~\cite{Herdeiro:2014goa,Herdeiro:2015gia}, a non-trivial distribution of a complex, massive scalar field can be added to Kerr BHs, keeping them asymptotically flat and regular on and outside the event horizon. Such a scalar field carries a conserved Noether charge, but which, unlike the electric charge, is not associated to a Gauss law. As such, it cannot be computed as a flux integral at infinity; it must be evaluated by a volume integral, summing up the appropriate component(s) of the conserved Noether current, from infinity up to the horizon. These solutions have an intimate connection to the superradiant instability of Kerr BHs~\cite{Herdeiro:2014ima}, in the presence of a massive scalar field (see~\cite{Brito:2015oca} for a review). They bifurcate from Kerr for particular backgrounds that can support a stationary scalar cloud in the linear theory~\cite{Hod:2012px,Hod:2013zza,Herdeiro:2014goa,Benone:2014ssa,Hod:2015ota,Hod:2015goa}, and reduce to boson stars~\cite{Schunck:2003kk}, horizonless gravitating solitons, when the horizon area vanishes. Kerr BHs with scalar hair (KBHsSH) can have phenomenological properties distinct from Kerr, for instance their shadows~\cite{Cunha:2015yba}. This fact, in view of the various observations/experiments that promise to deliver detailed information on BH candidates and strong gravity in the near future, makes their analysis in the astrophysical context quite timely -- see~\cite{Vincent:2016sjq,Ni:2016rhz} for recent examples of such phenomenological studies.

\bigskip

It is expectable that KBHsSH, just like the Kerr solution, admit electrically charged generalizations. Again, the astrophysical interest of such solutions is, perhaps, more limited, but understanding their existence and their physical properties is of relevance to fully grasp the impact of this scalar (or other) hair on the paradigmatic BHs of General Relativity. The purpose of this paper is, precisely, to construct examples of such electrically charged generalizations of KBHsSH and to examine some of their physical properties. 

\bigskip

We shall focus on Einstein--Maxwell--(complex)Klein-Gordon theory, where the scalar field is massive and has no self-interactions. All couplings, moreover, are minimal. We start by considering an ungauged (hence electrically uncharged) scalar field. In this case, the family of solutions -- Kerr-Newman BHs with ungauged scalar hair (KNBHsUSH) -- is described by four continuous parameters (with one non-trivial constraint between them): $(1)$ the ADM mass, $M$, which can be split into the horizon and exterior matter/energy contribution (composed of the scalar $\Psi$ plus electromagnetic fields), $M=M_{\rm H}+M^\Psi+M^{\rm EM}$; $(2)$ the total angular momentum, $J$, which can also be split in a similar fashion, $J=J_{\rm H}+J^\Psi+J^{\rm EM}$; $(3)$ the  Noether charge, $Q$, associated to the global $U(1)$ invariance of the complex scalar field, which obeys $Q=J^\Psi/m$, where $m$ is the azimuthal winding number; $(4)$ and the total electric charge, $Q_E$. All electric charge is contained within the BH horizon, $Q_E=Q_E^{\rm H}$, whereas all Noether charge is contained outside the horizon. By Gauss's law the former can be computed by the flux of the electric field on any closed 2-surface surrounding the horizon, and is unaffected by $Q$. The magnetic dipole moment, on the other hand, which is induced by the electric charge in the rotating spacetime, is affected by the Noether charge, with respect  to the KN value. This is appropriately described by the gyromagnetic ratio, which is $g=2$ for KN and it is $g\leqslant 2$ for the hairy BHs. Thus, for the same amount of total mass, angular momentum and electric charge, KNBHsUSH have less magnetic dipole moment, a suppression of the magnetic dipole due to the scalar hair. The domain of existence of KNBHsUSH is bounded by (a particular set of) KN BHs, when $Q=0$; a set of extremal (zero temperature) BHs; and (electrically uncharged) rotating boson stars, when $Q_E=0$, for which $M=M^\Psi$ and $J=J^\Psi=mQ$.  As for the uncharged KBHsSH, there is no static limit for KNBHsUSH.

\bigskip

Gauging the scalar field, with a gauge coupling $q_E$, hence endowing the scalar particles with electric charge, leads to a family of KN BHs with gauged scalar hair (KNBHsGSH), which exhibits some changes with respect to the previously discussed KNBHsUSH. 
Firstly, as in the ungauged case,
both the mass and the angular momentum can be split  into the horizon and exterior contribution.
However, the corresponding parts for the electromagnetic and  the scalar $\Psi$ fields cannot be rigorously
separated since  here they interact directly, not only via the spacetime geometry.
Moreover this time  
the total electric charge has both a horizon and a contribution sourced by the scalar field,  $Q=Q^{\rm H}_E+Q^\Psi_E$,  
and needs to be calculated by the asymptotic flux.
$Q_E^\Psi$ is determined by the total Noether charge, which counts the number of the scalar particles, multiplied by their individual electric charge:  $Q_E^\Psi=q_E Q/(4 \pi)$. 
This redistribution of the electric charge, does not, however change the qualitative behaviour of the gyromagnetic ratio. 
For all solutions constructed so far we still observe that it is always $g\leqslant 2$, with equality attained for the KN case. 
The boundaries of the domain of existence of KNBHsGSH are sensitive to the gauging. 
Both the KN, extremal and solitonic limits are different. 
In particular, the latter is a set of rotating  boson stars with  $nonzero$ electric 
and magnetic fields, for which  $Q_E= q_E Q/(4\pi)= q_E J/(4 \pi m)$.

\bigskip

This paper is organized as follows. In Section~\ref{sec_mod_u} we describe the ungauged scalar field model, the boundary conditions, physical quantities of interest and the numerical results for the domain of existence of the solutions, as well as some physical properties, including the gyromagnetic ratio. A similar, albeit less extensive analysis, is performed in Section~\ref{sec_mod_g} for the gauged case, emphasizing the differences with respect to the ungauged one. Finally, in Section~\ref{sec_remarks} we present some closing remarks.

\section{The ungauged scalar field model}
\label{sec_mod_u}

\subsection{Action, equations of motion and ansatz}
\label{sec_mofrl}
We start by considering Einstein--Maxwell theory, minimally coupled to a complex, massive (mass $\mu$)  ungauged scalar field $\Psi$.  The corresponding action is
\begin{eqnarray}
  \label{action}
 \mathcal{S} = \frac{1}{4\pi G}\int d^4x \sqrt{-g}\left[\frac{R}{4}- \frac{1}{4}F_{ab}F^{ab}- g^{ab}\Psi^*_{,a}\Psi_{,b} -\mu^2\Psi^*\Psi \right]\ ,  
\end{eqnarray}  
where $F_{ab}$ are the components of the Maxwell 2-form, $F$, related to the 1-form potential $A=A_adx^a$ as $F=dA$. The Einstein--Klein-Gordon--Maxwell equations, obtained by varying the action with respect to the metric, scalar field and electromagnetic field, are, respectively,
\begin{equation}
G_{ab}  = 2\left( T_{ab}^\Psi+T_{ab} ^{\rm EM} \right)\ , \qquad \Box \Psi =\mu^2\Psi \ , \qquad D_aF^a_{~b}=0 \ ,
\label{eom}
\end{equation}
where the two components of the energy-momentum tensor are
\begin{equation}
\label{emt}
T_{ab}^\Psi \equiv  
 \Psi_{ , a}^*\Psi_{,b}
+\Psi_{,b}^*\Psi_{,a} 
-g_{ab}  \left[ \frac{1}{2} g^{cd} 
 ( \Psi_{,c}^*\Psi_{,d}+
\Psi_{,d}^*\Psi_{,c} )+\mu^2 \Psi^*\Psi\right] \ , \qquad
 T_{ab}^{\rm EM} \equiv F_a^{~c}F_{bc} - \frac{1}{4}g_{ab}F_{cd}F^{cd} \ .
\end{equation}
This model is invariant under a \textit{global} transformation $\Psi\rightarrow \Psi e^{i\alpha}$, where $\alpha$ is constant.

KNBHsUSH are obtained using the metric, scalar field and electromagnetic potential ansatz given by
\begin{align}
  \label{metric_ansatz}
  ds^2 &= -e^{2F_0}Ndt^2 + e^{2F_1}\left( \frac{dr^2}{N} + r^2d\theta^2 \right) + e^{2F_2}r^2\sin^2\theta \left(d\varphi - Wdt \right)^2 \ ,\\
\label{scalar_ansatz}
\Psi &= \phi(r,\theta)e^{i(m\varphi-w t)}~, \\
 \label{electric_ansatz}
 A_adx^a &= \left( A_t - A_\varphi W\sin\theta \right)dt + A_\varphi\sin\theta d\varphi \ ,
\end{align}
where $N\equiv 1-r_H/r$ and $r_H$ is a constant describing the event horizon location in this coordinate system; the metric ansatz functions $F_i,W$, $i=0,1,2$, as well as $\phi$, $A_t$ and $A_\varphi$, depend on the spheroidal coordinates $r$ and $\theta$ only; $w\in \mathbb{R}^+$ is the scalar field frequency and $m=\pm 1,\pm 2$\dots is the azimuthal harmonic index. In the following we shall focus on the case $m=1$ as an illustrative set of solutions, and also nodeless solutions for the scalar field profile $\phi$. Solutions with nodes will also exist, corresponding to excited states with higher ADM mass.

\subsection{Boundary conditions}
\label{sec_bc}
In order to find KNBHsUSH, we use the numerical strategy and code already discussed in some of our previous works (see, $e.g.$~\cite{Herdeiro:2015gia,Herdeiro:2016tmi}). To obtain these solutions, appropriate boundary conditions must be imposed, that we now summarize.
  
At spatial infinity, $r\rightarrow\infty$, we require that the solutions approach a Minkowski spacetime
with vanishing matter fields:
\begin{equation}
  \lim_{r\rightarrow \infty}{F_i}=\lim_{r\rightarrow \infty}{W}=\lim_{r\rightarrow \infty}{\phi}=\lim_{r\rightarrow\infty}A_\varphi=\lim_{r\rightarrow\infty}A_t=0\ .
\end{equation}
(Observe that the last equality could be changed to a constant, rather than zero, in a different gauge.) 
%
On the symmetry axis, $i.e.$ at $\theta=0,\pi$, axial symmetry and regularity require that
\begin{equation}
\partial_\theta F_i = \partial_\theta W = \partial_\theta A_t = \phi = A_\varphi = 0\ .
\end{equation} 
Moreover, all solutions herein are invariant under a reflection in the equatorial plane ($\theta=\pi/2$).
The event horizon is located at a surface with constant radial variable, $r=r_H>0$.
By introducing a new radial coordinate $x=\sqrt{r^2-r_H^2}$ 
the boundary conditions and numerical treatment of the problem are simplified.
Then one can write an approximate form of the solution  near the horizon as a power series in $x$,
which implies the following boundary conditions
\begin{equation}
\partial_x F_i \big|_{x=0}= \partial_x \phi  \big|_{x=0} =  0,~~W \big|_{x=0}=\Omega_H,~~
A_t \big|_{x=0} =  \Phi_H,~~ \partial_x A_\varphi \big|_{x=0}=0 ,
\label{bch1}
\end{equation}
where $\Omega_H $ is the horizon angular velocity
and $\Phi_H$ is the horizon electrostatic potential. 
Similarly to the uncharged case, the existence of a smooth solution imposes also the
\textit{synchronization condition}
\begin{eqnarray}
\label{cond}
w=m\Omega_H\ .
\end{eqnarray}
%

\subsection{Physical quantities}
\label{sec_pq}
Axi-symmetry and stationarity of the spacetime (\ref{metric_ansatz})
 guarantee the existence of two conserved global charges, the total mass $M$ and angular momentum $J$, 
which can be computed either as Komar integrals at spatial infinity or, equivalently, 
from the decay of the appropriate metric functions:
\begin{eqnarray}
\label{asym}
g_{tt} =-e^{2F_0}N+e^{2F_2}W^2r^2 \sin^2 \theta 
\to
 -1+\frac{2GM}{r}+\dots, ~~
g_{\varphi t}=-e^{2F_2}W r^2 \sin^2 \theta
\to 
\frac{2GJ}{r}\sin^2\theta+\nonumber \dots.  
\end{eqnarray}
These quantities can be split into the horizon contribution -- computed as a Komar integral on the horizon -- and the matter contributions, composed of the scalar field and electromagnetic parts, computed as the volume integrals of the appropriate energy-momentum tensor components: 
\begin{eqnarray}
\label{MH-hor}
M=M^\Psi+M^{\rm EM}+M_H\ , \qquad J=J^\Psi+J^{\rm EM}+J_H\ ,
\end{eqnarray}
where $M_H$ and $J_H$  are the horizon mass and angular momentum.
$M^\Psi$ and $J^\Psi$ are the scalar field energy and angular momentum outside the horizon,
with  
\begin{align}
\label{Mpsi}
-M^\Psi\equiv  \int_{\Sigma} dS^a (2T_{ab}^\Psi \xi^b-T^\Psi\xi_a)
 = 4\pi \int_{r_H}^\infty dr \int_0^\pi d\theta~r^2\sin \theta ~e^{F_0+2F_1+F_2}
 \left(
 \mu^2-2 e^{-2F_0}\frac{w(w-mW)}{N}
 \right)\phi^2 ~~,
\end{align}
while $J^\Psi=mQ$,
where $Q$ is the Noether charge 
associated with the the $global$ $U(1)$ symmetry of the complex scalar field 
\begin{eqnarray}
\label{Q-int}
Q=4\pi \int_{r_H}^\infty dr \int_0^\pi d\theta 
~r^2\sin \theta ~e^{-F_0+2F_1+F_2}  \frac{(w-mW)}{N}\phi^2 ~.
\end{eqnarray}
To measure the hairiness of a BH, it is convenient 
to introduce the normalized Noether charge
\begin{eqnarray}
\label{q}
q=\frac{mQ}{J}~,
\end{eqnarray}
with $q=1$ for solitons and $q=0$ for KN BHs.
Also, $M^{\rm EM}$ and $J^{\rm EM}$ are the mass and angular momentum stored in the electromagnetic field
outside the horizon.  
The solutions possess also an electric charge $Q_E$ that can be computed using Gauss's law, 
on any closed 2-surface covering the horizon. 
Alternatively, $Q_E$ can be computed from the decay of the 4-potential, together with the magnetic dipole moment $\mu_M$:
 \begin{eqnarray}
 \label{asym-matter-fields}
A_t\sim 
\frac{Q_E}{r}+\dots \ , \qquad A_{\varphi}\sim \frac{\mu_M \sin \theta}{r}+\dots\
 .
 \end{eqnarray}
As with the KN BHs, the gyromagnetic ratio $g$ defines how the magnetic dipole moment 
is induced by the total angular momentum and charge, for a given total mass:
 \begin{equation}
 \mu_M=g\frac{Q_E}{2M}J \ .
 \label{gyro}
 \end{equation}
The BH horizon introduces a temperature $T_H$ and an entropy $S={A_H}/({4G})$,
where 
\begin{eqnarray}
\label{THAH}
T_H=\frac{1}{4\pi r_H}e^{(F_0-F_1)|_{r_H}}\ ,
\qquad
A_H=2\pi r_H^2 \int_0^\pi d\theta \sin \theta~e^{(F_1+F_2)|_{r_H}} \ .
\end{eqnarray}
The various quantities above are related by a Smarr mass formula 
\begin{eqnarray}
\label{smarr}
M=2 T_H S +2\Omega_H (J-m Q) + \Phi_H Q_E+ M^\Psi,
\end{eqnarray}
The solutions satisfy also the 1st law 
\begin{eqnarray}
\label{first-law}
dM=T_H dS +\Omega_H dJ + \Phi_H dQ_E\ .
\end{eqnarray}
Finally, observe that \eqref{smarr} and \eqref{MH-hor} are consistent with a different Smarr relation, only in terms of horizon quantities  
\begin{eqnarray} 
\label{rel-hor}
M_H=2T_H S+2 \Omega_H J_H~,
\end{eqnarray}
together with the electromagnetic relation $M^{\rm EM}=\Phi_HQ_E+2\Omega_HJ^{\rm EM}$.

\subsection{The results }
\label{sec_results_u}
As in the previous work \cite{Herdeiro:2015gia,Herdeiro:2016tmi}
the numerical integration is performed with 
dimensionless variables introduced by using natural units set by $\mu$ and $G$.
The global charges and all other quantities of interest are also
 expressed in units set by $\mu$ and $G$ 
(we set $G=1$ in what follows). 
In particular this means we set $\mu Q_E\rightarrow Q_E$; 
note that $\Phi_H$ is dimensionless (in units such that $4\pi \epsilon_0=1$). 

Let us start by getting an overview of the domain of existence of KNBHsUSH. 
This requires fixing the new degree of freedom, related to electric charge. 
An important observation here is that, similarly to the
KN case, no solitonic limit exists, for a nonzero $Q_E$.
Thus, for most of the numerical solutions we have chosen to fix the electrostatic potential on the horizon $\Phi_H$
and vary the remaining  input parameters $\Omega_H$ and $r_H$.
This allows these  two-dimensional sections of the full domain of existence to reach the solitonic limit, 
wherein the horizon area vanishes and the electrostatic potential becomes constant everywhere and pure gauge.

\begin{figure}[h!]
  \begin{center}
    \includegraphics[width=0.497\textwidth]{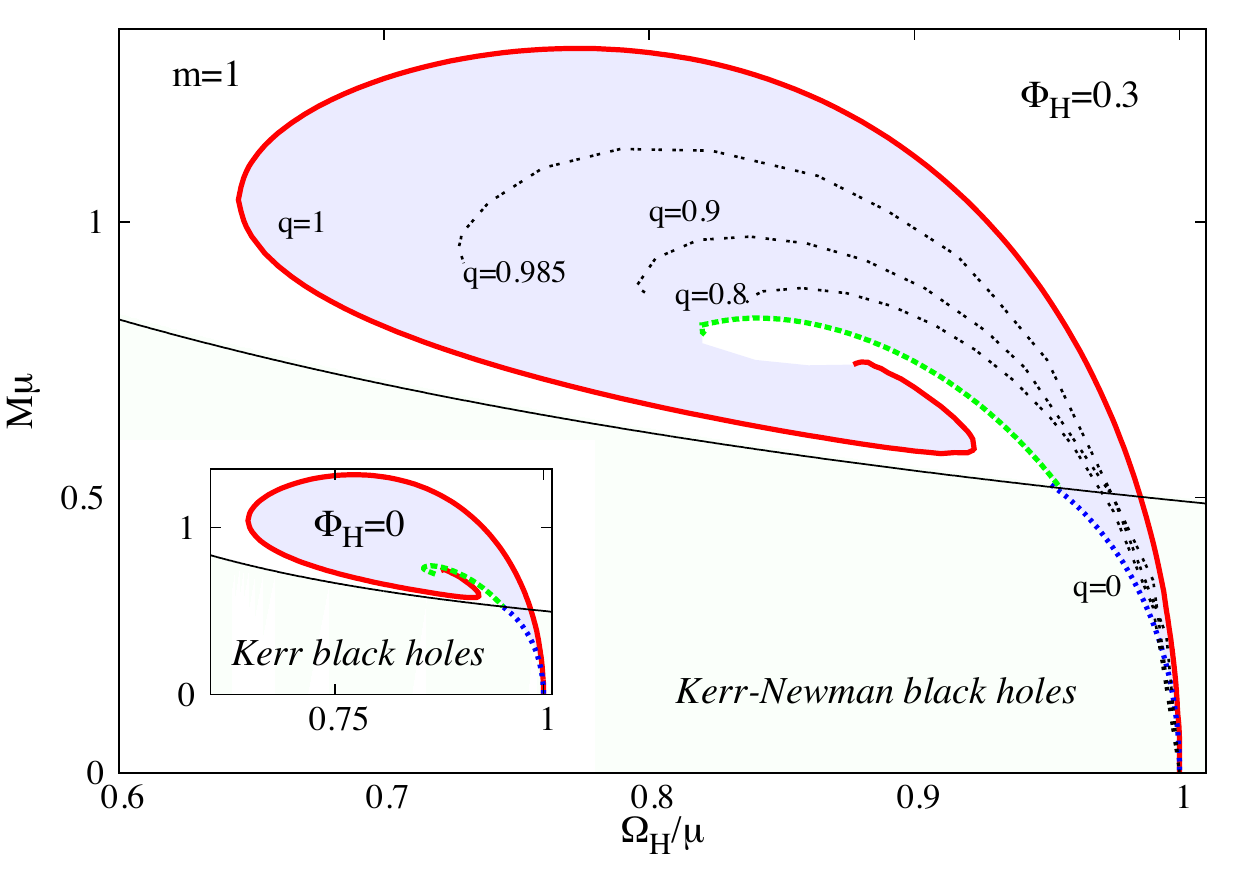}
     \includegraphics[width=0.497\textwidth]{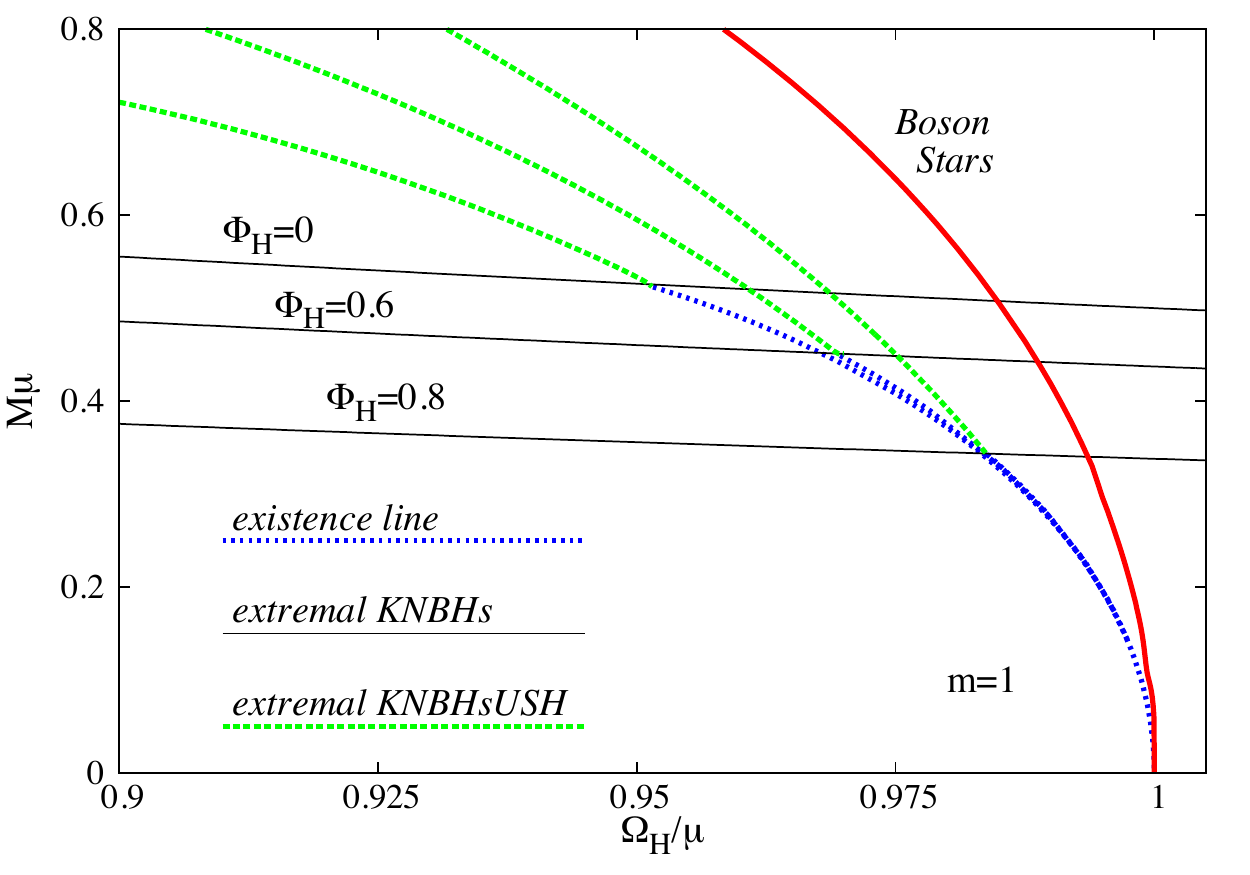}
  \end{center}
  \caption{The $(\Omega_H,M)$ domain of existence for a sample of KNBHsUSH. (Main left panel) Diagram for $\Phi_H=0.3$ with the boson star envelope (red solid line), the existence line on the domain of KN BHs (blue dotted line) and the line of extremal KNBHsUSH (green dashed line). The black solid line corresponds to the extremal KN BHs; non-extremal solutions exist below. The black dotted lines have constant normalized Noether charge $q$. (Inset) diagram for $\Phi_H=0$, for comparison. (Right panel) Detail around the intersection of the existence lines with the extremal KNBHsUSH lines and the extremal KN lines for $\Phi_H=0,0.6$ and $0.8$.  
	}
  \label{fig:w-M}
\end{figure}

 In Fig.~\ref{fig:w-M} (left panel), 
we exhibit the $(\Omega_H,M)$ domain of existence of the solutions, where we have fixed the horizon electrostatic potential to be $\Phi_H=0.3$. Observe that the domain therein was obtained by extrapolating into the continuum
the results from discrete sets of around two thousand numerical solutions; we also remark that a qualitatively similar picture has been found for $\Phi_H=0.6$.
 As shown in the main panel (the inset is for $\Phi_H=0$), 
this domain of existence is bounded by boson stars (red solid line), 
the KN limit (blue dotted line -- dubbed \textit{existence line}) 
and the extremal KNBHsUSH limit (green dashed line). 
The last two limits vary with the electrostatic potential whereas the first one does not; 
this can be observed in the right panel, where part of the line of extremal KNBHsUSH is shown for $\Phi_H=0; 0.6$ and $0.8$, as well as the corresponding existence line and the line for extremal KN BHs (black solid lines). 
The trend is that the larger the electrostatic potential becomes, the lower the mass of the extremal KN BH is, along the existence line (henceforth dubbed as \textit{Hod point}, following~\cite{Herdeiro:2015tia}), whence the line of extremal KNBHsUSH starts. This is the expected result from the known behaviour of KN BHs. Another trend, illustrated by comparing the main left panel with the inset, is that for higher $\Phi_H$, there are extremal hairy BHs with lower horizon angular velocity.

 In Fig.~\ref{fig:w-g} (left panel), we exhibit the ratio $ M^{\Psi}/M$, which gives another measure of hairiness
   as a function of $\Omega_H$,
for $\Phi_H=0.3$. The figure shows that small fractions of the total energy in the hair are only allowed for sufficiently large horizon angular velocity. When the angular velocity is small, equilibrium between the hair and the horizon is only possible for solutions with $q$ close to unity, $i.e.$, boson star-like. 
The inset in this figure shows the $(\Omega_H,Q_E)$ domain of existence of the KNBHsUSH solutions. It illustrates that the electric charge of the solutions, for fixed $\Omega_H$ between that of the Hod point and the maximum allowed frequency, $\Omega_H=\mu$, is maximized along the existence line (and in particular at the Hod point). But for lower values of the frequency, slightly larger charges than that found at the Hod point are possible, occurring along the extremal hairy BHs line.

\subsubsection{Gyromagnetic ratio}
Rotating charges give rise to a magnetic dipole moment, $\mu_M$. 
In classical electromagnetism, a generic relation of the form~\eqref{gyro}, 
between $\mu_M$ and the total angular momentum, mass and charge can be derived, 
for systems with constant ratio of charge to mass density, yielding $g=1$. 
When experiments such as that performed for Stern and Gerlach in the early XXth century, 
showed that the electron should have $g=2$, it became clear that a new fundamental description 
for the electron was necessary, beyond the scope of the non-relativistic quantum theory. 
Such a description appeared with the Dirac equation, which, from first principles predicts $g=2$, 
a value that is corrected by loop diagrams in Quantum Electrodynamics (QED), 
yielding the so called anomalous magnetic moment, whose agreement 
with experiment is one of the outstanding successes of QED.

In BH physics, Carter first observed that $g=2$ for the KN solution. 
Since then many other studies considered the gyromagnetic ratio of rotating charged BHs, 
for instance with different asymptotics and in higher dimensions (see $e.g.$~
\cite{Garfinkle:1990ib,Herdeiro:2000ap,Aliev:2004ec,Ortaggio:2006ng,Aliev:2006tt}). 
Here we show that the addition of scalar hair leads to a suppression of the gyromagnetic ratio, 
and of the corresponding magnetic dipole moment, 
with respect to that of a comparable KN BH.  
 A novel aspect is that $g$ 
can be smaller than 1, a rather unsual feature in other models of relativistic, 
charged and spinning compact objects, 
$cf.$~\cite{Novak:2003uj}.

\begin{figure}[H]
  \begin{center}
    \includegraphics[width=0.497\textwidth]{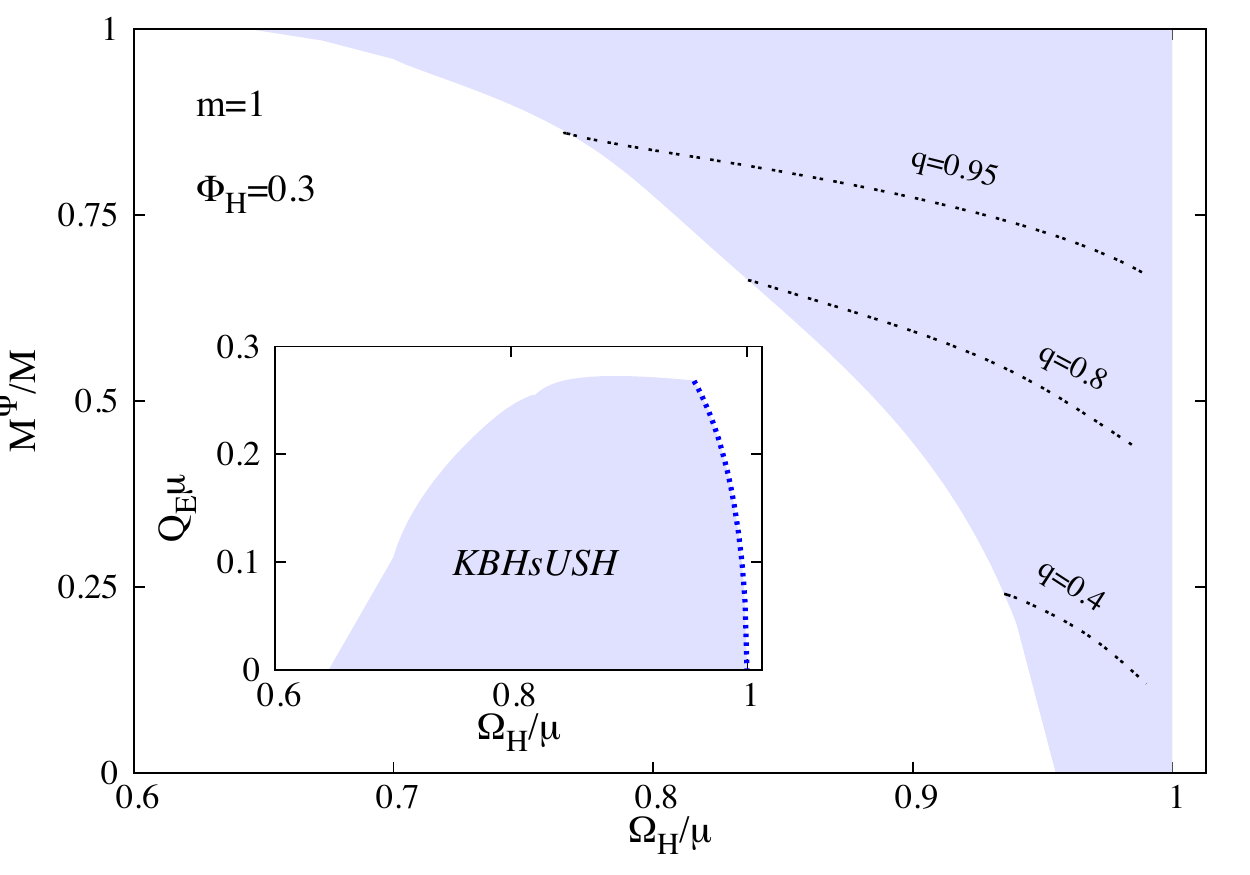}
      \includegraphics[width=0.497\textwidth]{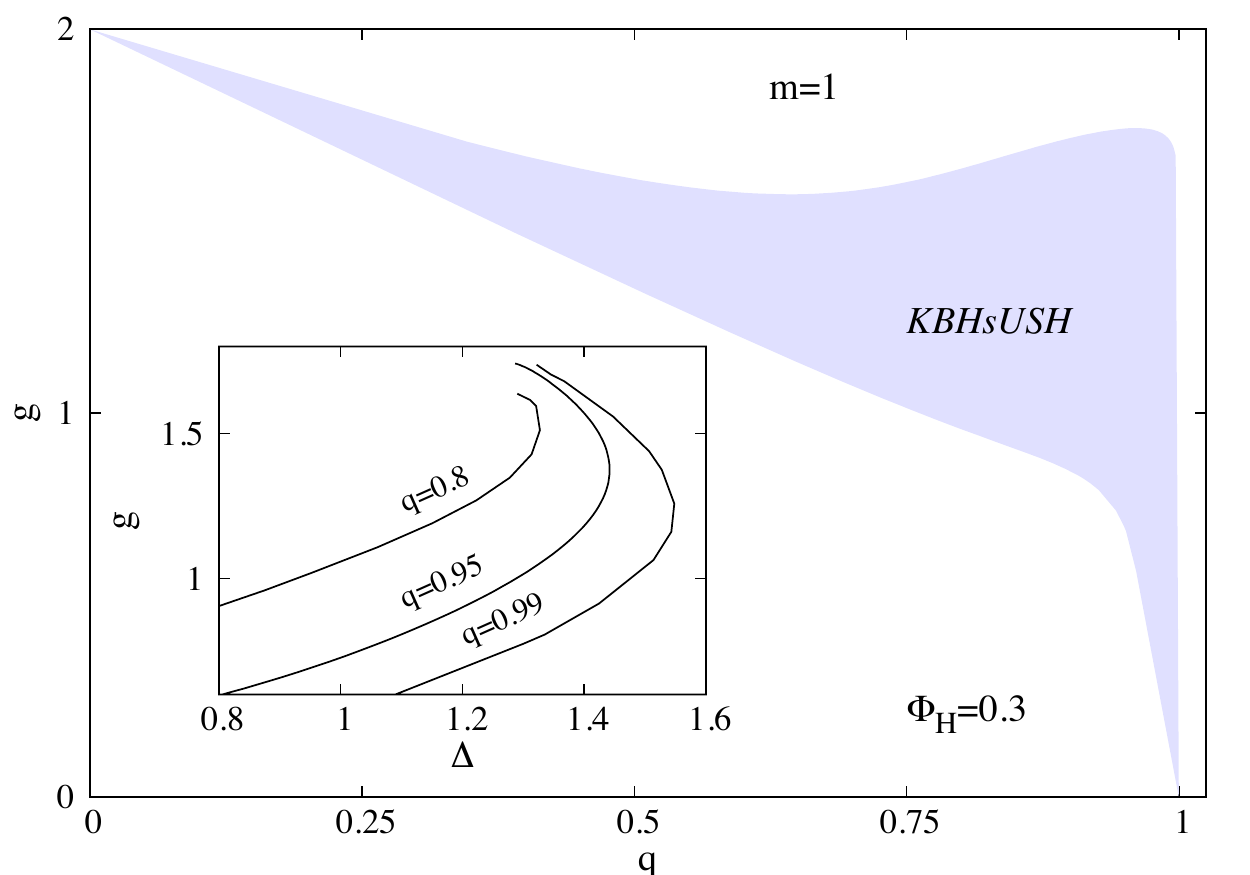}
  \end{center}
  \caption{(Left panel) The ratio $M^\Psi/M$ is shown as a function of $\Omega_H$ for a sample of KNBHsUSH. The inset shows the electric charge as a function of $\Omega_H$, where the blue dotted line is the existence line. 
	(Right panel)  The $(q,g)$ space. The inset show $g$ as a function of $\Delta$, that determines the KN bound.
}
 \label{fig:w-g} 
\end{figure}

In Fig.~\ref{fig:w-g} (right panel), we exhibit the gyromagnetic ratio in a $(q,g)$-diagram, for KNBHsUSH with $\Phi_H=0.3$.
The diagram shows that the gyromagnetic ratio, $g$, of both the extremal and non-extremal hairy BH solutions, is always less than $2$.
As expected, it does approach $2$, for both cases, in the limit of vanishing hair.
Further insight is obtained by considering the quantity
\begin{eqnarray}
\Delta\equiv \frac{M^2}{Q_E^2+J^2/M^2} \ ,
\end{eqnarray}
which determines the KN bound $\Delta\geqslant 1$. Indeed, all KN BHs have $\Delta>1$. This bound is, however, violated by a large set of KNBHsUSH, in particular by those close to the BS limit. This is reminiscent of what has been found for KBHsSH - see the discussions in~\cite{Herdeiro:2014goa,Herdeiro:2015gia,Herdeiro:2015tia,Delgado:2016zxv}.  Our results show that
solutions with $g<1$ predominantly exhibit $\Delta<1$ and thus violate the KN bound -- $cf.$ the inset of Fig.~\ref{fig:w-g} (right panel).

\section{Gauged scalar field model}
\label{sec_mod_g}

\subsection{Main differences in the model}


Let us now consider the model described in Section~\ref{sec_mod_u} but with a \textit{gauged} scalar field, 
that couples minimally to the electromagnetic field, with gauge coupling $q_E$. 
This coupling is implemented by replacing the partial derivatives of the scalar field in the action~\eqref{action} as
\begin{equation}
\partial_a \Psi \longrightarrow D_a\Psi=\partial_a \Psi + iq_E A_a \Psi \ .
\label{mc}
\end{equation}
The Einstein equations still take the form~\eqref{eom}, but with the substitution~\eqref{mc} 
in the scalar field energy-momentum tensor~\eqref{emt}. Then, the scalar and Maxwell equations of motion become
\begin{eqnarray}
\label{field-eqs}
D_{a}D^{a}\Psi=\mu^2 \Psi\ , \qquad 
\nabla_{b}F^{ba}=
iq_E \big [ (D^{a}\Psi^*) \Psi-\Psi^*(D^a \Psi) \big ] 
\equiv q_E j^a  \ .
\end{eqnarray}  
Physically, the scalar field is now electrically charged 
(its quanta, the scalar particles, carry a charge $q_E$), and thus the scalar field sources the Maxwell field.

This model is invariant under the $local$ U(1) gauge transformation 
\begin{eqnarray}
\label{gauge-transf}
\Psi \to \Psi e^{-i q_E \alpha}\ ,~~A_a\to A_a+\partial_a \alpha \ ,
\end{eqnarray}
where $\alpha$ is a real function. One consequence of this gauge invariance is that the $(t, \varphi)$-dependence of the scalar field ansatz~\eqref{scalar_ansatz}, 
can now be gauged away by applying the local $U(1)$ symmetry
(\ref{gauge-transf})
with $\alpha =  (m\varphi -w t)/q_E$.  This, however, also changes the gauge field, as $A_t\to A_t-w/q_E$,
$A_\varphi \to A_\varphi+m/q_E$. 
Consequently, the solutions cannot be constructed starting with the configurations in the
previous sections and increasing $q_E$.
Thus, in order
to be able to consider this approach, we  keep the $(t,\varphi)$-dependence in 
the scalar field ansatz and fix the corresponding gauge freedom by setting $A_t = A_\varphi = 0$ at infinity.

One major difference with respect to the case 
discussed in the previous section 
 is that the solitonic limit of the solutions carries now a nonzero electric charge.
Self-gravitating charged boson stars were first considered, in spherical symmetry, in~\cite{Jetzer:1989av} 
(see also the recent work  \cite{Pugliese:2013gsa}). 
To the best of our knowledge, no rotating generalizations of these static solutions have been reported\footnote{ 
Some properties of the spinning charged solitons, 
with a self-interacting ($Q$-ball type) scalar field model, were addressed in~\cite{Brihaye:2009dx}.
}.
The Noether charge $Q$ of the solitons, $i.e.$ the total
particle number, is now intrinsically related to the electric charge $Q_E$. 
The former can be computed as  
\begin{eqnarray}
\label{Q1}
Q= \int j^t \sqrt{-g} dr  d\theta d\varphi=
 4\pi \int_{0}^\infty dr \int_0^\pi d\theta  
~r^2\sin \theta ~e^{-F_0+2F_1+F_2}  (w-q_E A_t -mW)\phi^2 \ ,
\end{eqnarray}
whereas the latter is read from the asymptotics
 of the electric potential $A_t$, as given in  (\ref{asym-matter-fields}). 
A straightforward computation 
shows that both the Noether charge and the electric charge of the spinning \textit{solitons}
are proportional  to the total angular momentum,
\begin{eqnarray}
\label{JQ}
J= m Q=\frac{4 \pi m Q_E}{q_E}\ .
\end{eqnarray}

\subsection{Features of the gauged scalar field solutions}
\label{sec_results_g}
 
The construction of the  gauged scalar field solutions is similar 
to that described above for the ungauged case ($q_E=0$ limit).
In particular, the KNBHsGSH are subject to the same set of 
boundary conditions  used in the ungauged case.
The synchronization condition, however, is different,   
\begin{equation}
\label{cond-new}
 w-q_E \Phi_H=m \Omega_H \ ,
\end{equation}
in agreement with the result found in the linear theory~\cite{Hod:2014baa,Benone:2014ssa}.

The electrically charged boson stars also form a part of the domain of existence of KNBHsGSH. 
Thus we have paid special attention to this limiting case.
These solutions  are obtained by considering the ansatz~\eqref{metric_ansatz}--\eqref{electric_ansatz} with $r_H=0$ 
and replacing the boundary conditions at the horizon~\eqref{bch1} 
by the following boundary conditions at the origin
\begin{eqnarray}
\label{bc0} 
\partial_r F_i|_{r=0}= 
W|_{r=0}=0\ ,~~
\phi| _{r =0}=0\ ,~~\partial_r A_t|_{r=0}=A_\varphi|_{r=0}=0\ .
\end{eqnarray}

\begin{figure}[H]
  \begin{center}
    \includegraphics[width=0.497\textwidth]{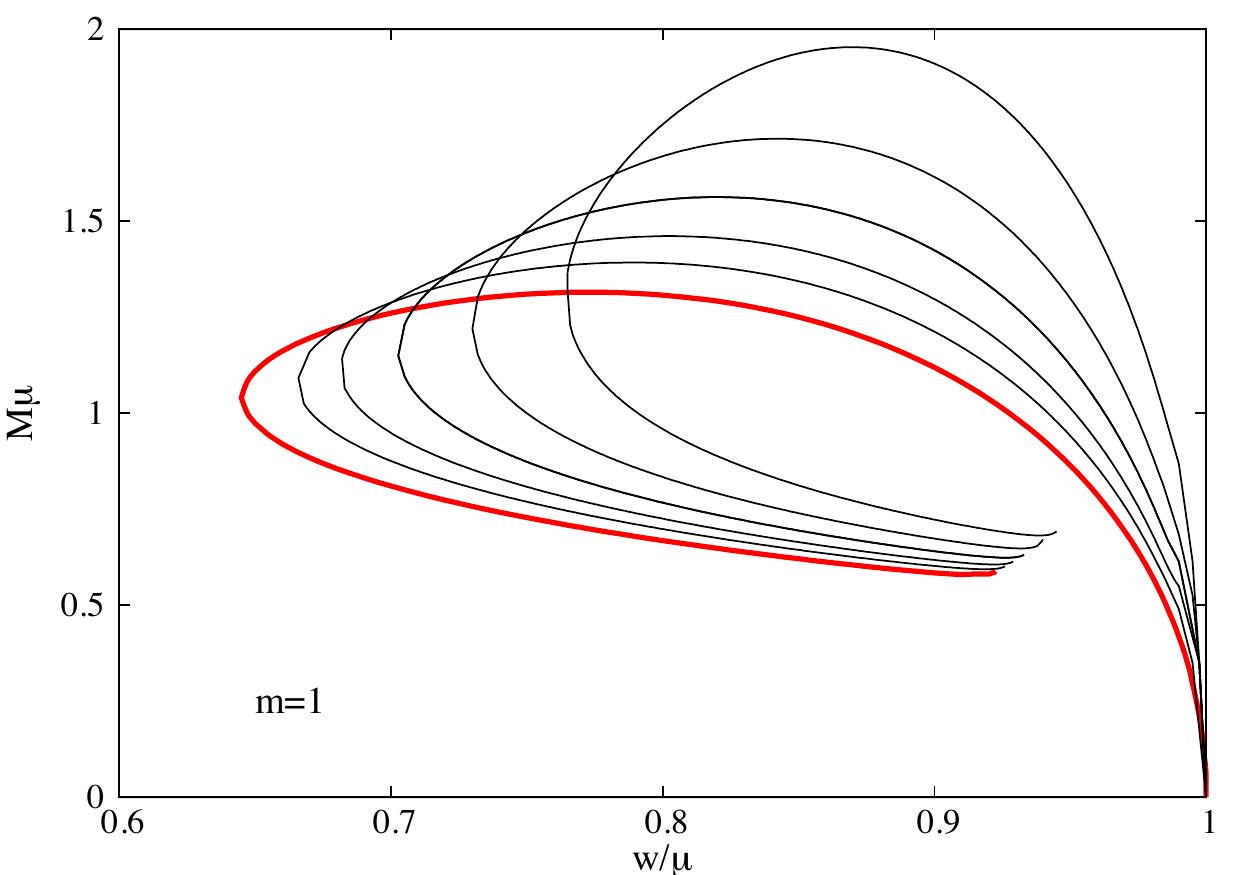}
      \includegraphics[width=0.497\textwidth]{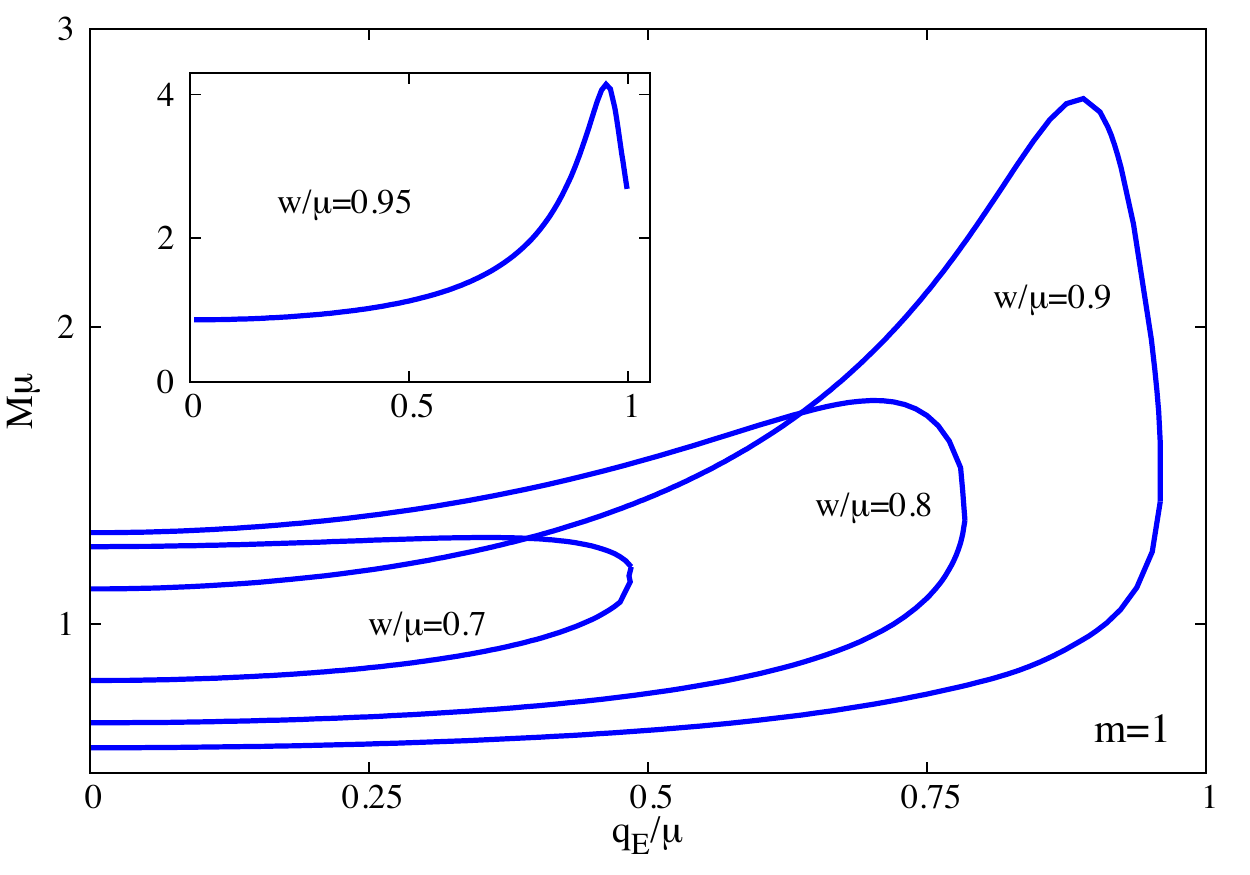}
  \end{center}
  \caption{
	(Left panel)
	The $(w,M)$ diagram for spinning boson stars with $q_E=0$ (red curve), $q_E/\mu=0.2, 0.3, 0.4, 0.5$
	and $0.6$ (top curve).  
		(Right panel)
	The mass $M$ is shown as a function of the gauge coupling constant $q_E$
	for several frequencies, $w/\mu=0.7, 0.8, 0.9$ and $0.95$ (as an inset).	}
  \label{fig:w-M-gauged}
\end{figure}

Some results of the numerical integration are shown in Fig.~\ref{fig:w-M-gauged} (left panel).
The basic properties of the spinning gauged boson stars solutions can be summarized as follows.
First, for all values of the gauge coupling considered so far, 
the frequency dependence of the solutions is qualitatively similar to the ungauged case.
The solutions 
exist for a limited range of frequencies
$0<w_{min}<w<\mu$. 
In particular, we observe that the minimal frequency increases with $q_E$. 
After this minimal frequency, a backbending towards larger values of $w$ occurs, yielding a second branch of solutions. 
A second backbending, towards
smaller values of $w$,  is observed as the frequency reaches a maximal value along the second branch, $w\to w_{max}$, whose value increases again with $q_E$. 
Then, similarly to the ungauged limit, a third branch of solutions develops -- not shown in Fig.~\ref{fig:w-M-gauged} (left panel).
Subsequently, we expect the existence of an inspiraling behaviour of the solutions, 
in analogy with uncharged boson stars, towards a limiting configuration.

Although only the mass is displayed in Fig.~\ref{fig:w-M-gauged} (left panel), the $J(w)$
diagram has a very similar shape.
Consequently,  the axially symmetric gauged boson stars do not possess
a static limit.  Observe also that the maximal mass of spinning gauged boson stars solutions increases with $q_E$.

As shown in Fig.~\ref{fig:w-M-gauged} (right panel), 
the solutions possess also a nontrivial dependence on 
the gauge coupling constant $q_E$.
For given values of $w$,
spinning solutions exist up to a maximal
value of the gauge coupling constant only, $q_E=(q_E)_{max}$.
The physics rationale behind this behaviour 
is similar to that discussed for the spherically symmetric case 
\cite{Jetzer:1989av,Pugliese:2013gsa}.
For $q_E>(q_E)_{max}$ the charge repulsion  becomes bigger than
  the scalar and gravitational attraction and localized solutions cease to exist  
(note that the maximal value of $q_E$
increases with frequency).
Also, as seen in Fig.~\ref{fig:w-M-gauged} (right panel),
all global charges stay finite as $q_E\to (q_E)_{max}$.

\begin{figure}[H]
  \begin{center}
    \includegraphics[width=0.497\textwidth]{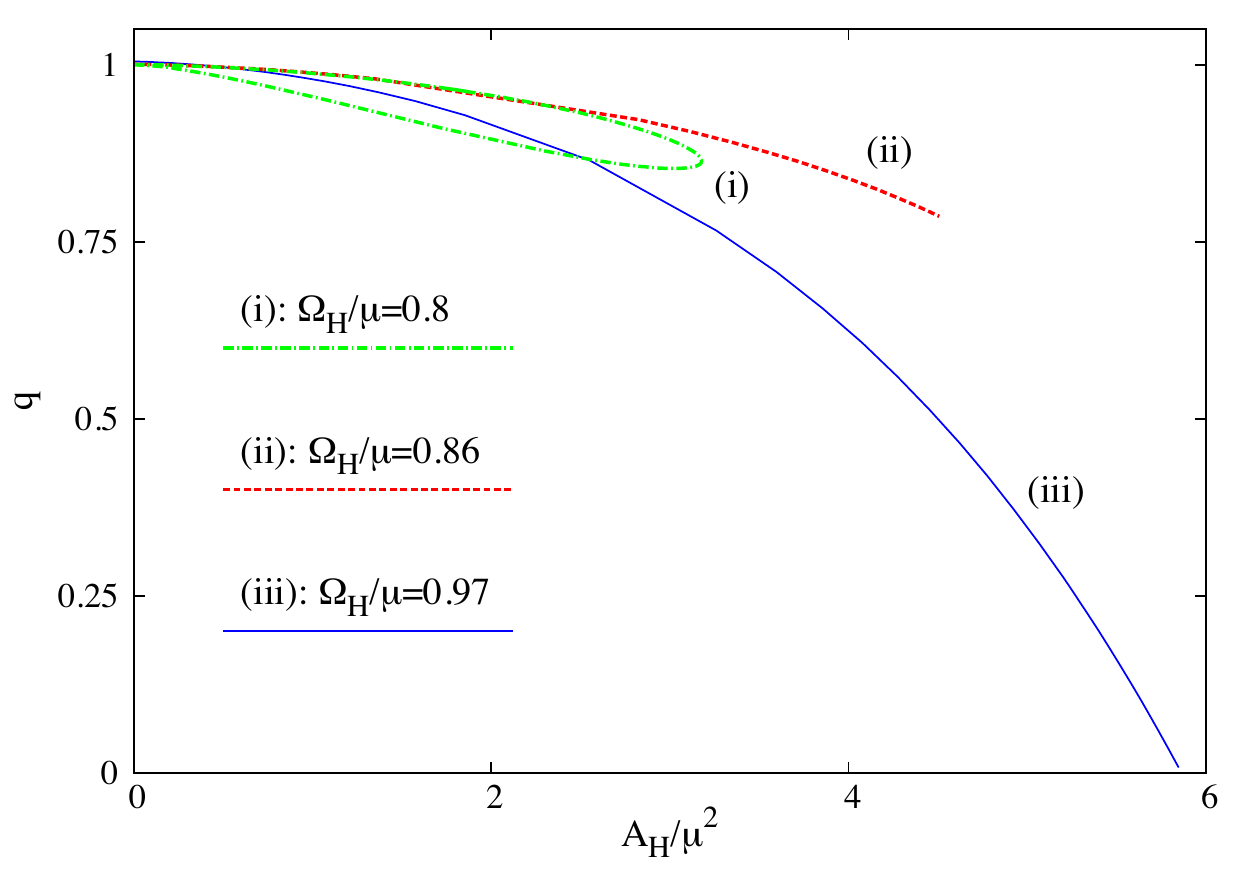} 
      \includegraphics[width=0.497\textwidth]{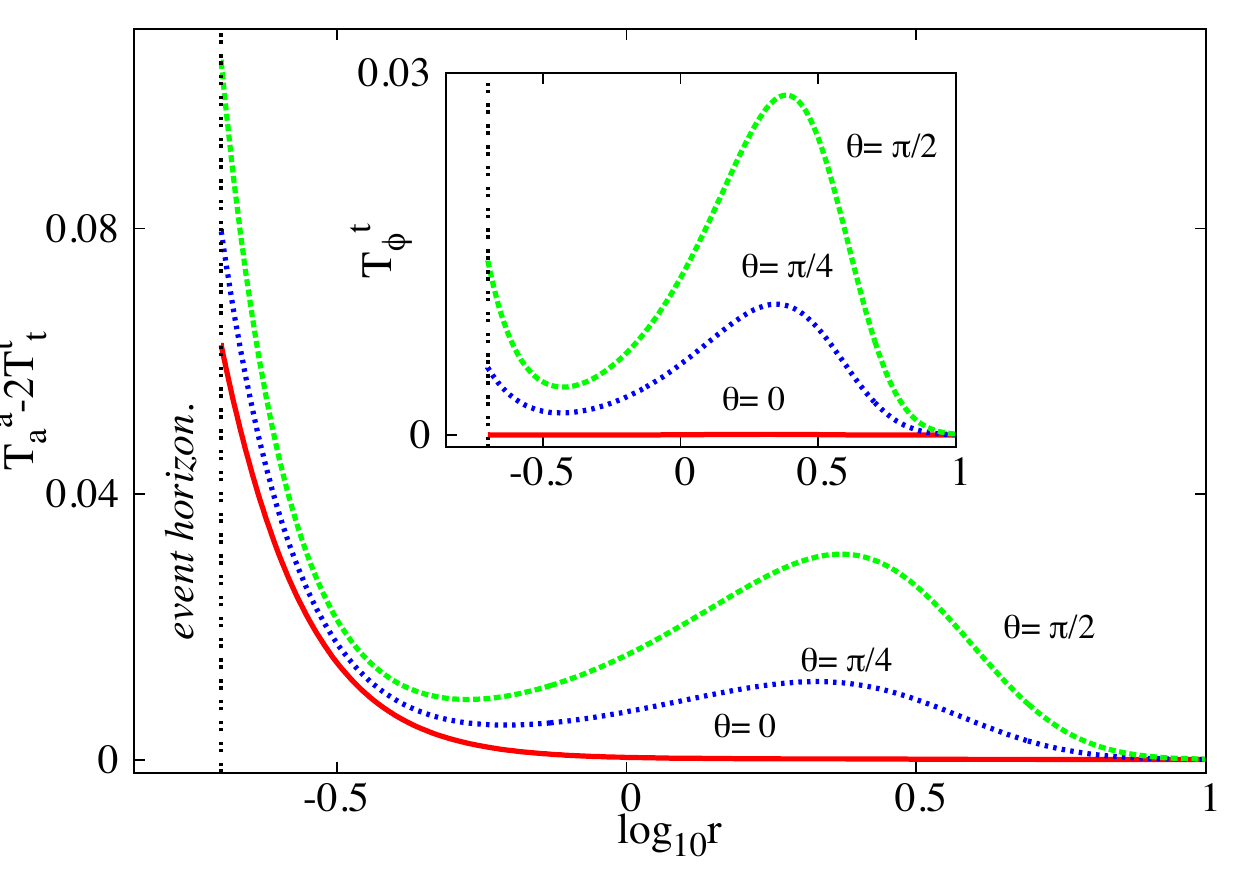} 
  \end{center}
  \caption{ (Left panel) 
	The $(A_H,q)$ diagram is shown for three sets of KNBHsGSH solutions with fixed values of 
$\Omega_H$ and 
$q_E/\mu=0.2$, 
$\Phi_H=0.1$. (Right panel) 
 Energy density (and angular momentum density in the inset) along three different slices of constant $\theta$ for an illustrative example of a KNBHGSH.
	}
  \label{fig:q-AH-gauged}
\end{figure}

KNBHsGSH are obtained by adding a horizon at the center of the spinning gauged boson star  
we have just described, which can be done for  any such  solution.
One way to construct the BHs
  is to  start from  boson stars 
and slightly increase the horizon size via the parameter $r_H$.
In this approach, the other input parameters 
$\Omega_H$, $q_E$,  $\Phi_H$ and $m$
are kept fixed. 
We recall that for BHs, the frequency $w$ is fixed by the synchronization condition (\ref{cond-new}).
Then one finds three 
possible behaviours for
the resulting branches of BH solutions -- see Fig.~\ref{fig:q-AH-gauged} (left panel).
 First  ($i$), for small enough values of $\Omega_H$,
 the branch of BHs connects two different boson stars; 
as $r_H\to 0$ the horizon area vanishes, $q\to 1$, while the temperature
 diverges.
For intermediate values of  $\Omega_H$,
the branch of solutions ends in an extremal KNBHsGSH solution ($ii$). 
These limiting configurations have finite
horizon size   and global charges, $0<q<1$ and appear to possess 
a regular horizon.
Finally   ($iii$), for large values of $\Omega_H$,
the branch of  KNBHsGSH interpolates between 
a charged boson star and a set of critical KN solutions (with $q=0$ and $A_H>0$), which lies again 
on an {\it existence line}. 

 In Fig.~\ref{fig:q-AH-gauged} (right panel) we exhibit the (Komar) energy density and angular momentum density (in the inset) for an illustrative example of a KNBHGSH  with physical input parameters 
$r_H=0.24, w =0.86, q_E =0.2, \Phi_H=0.1$. 
These densities have a contribution from both the electromagnetic and the scalar field. The main feature we wish to emphasize is the composite structure revealed by the plots. KN BHs have an (electromagnetic) energy and angular momentum density that decays with the radial coordinate, whereas KBHsSH 
(and boson stars)
have toroidal-like distributions for the (scalar) energy and angular momentum densities. 
Consequently, KNBHsGSH exhibit a superposition of these two behaviours, with decaying densities from the horizon but which exhibit a local maximum, in the neighbourhood of the equatorial plane, at some finite radial coordinate. 
A similar energy and angular momentum distribution can be found for KNBHsUSH.  

The behaviours illustrated in Fig.~\ref{fig:q-AH-gauged} supports the expectation that the domain of existence of KNBHsGSH will fill in the domain delimited by the boson star curves exhibited in Fig.~\ref{fig:w-M-gauged} (left panel), together with the existence line of KN BHs and a line of extremal KNBHsGSH,  in a qualitatively similar fashion to that shown in Fig.~\ref{fig:w-M} (left panel). Having established these solutions exist, and that their domain of existence will be analogue to the case of KNBHsUSH, we will not enter further details here; its systematic and detailed study will be reported elsewhere.
Here we mention only that, similar to the ungauged case, 
the gyromagnetic ratio of  KNBHsGSH constructed so far is always smaller than $g=2$.

\section{Remarks}
\label{sec_remarks}

The Kerr solution~\cite{Kerr:1963ud}, which describes the paradigmatic BH geometry in General Relativity, allows a generalization with electric (or magnetic) charge~\cite{Newman:1965my}, discovered shortly after the Kerr metric. Much more recently, it was found that the Kerr solution also allows a generalizations with scalar~\cite{Herdeiro:2014goa,Herdeiro:2015gia,Kleihaus:2015iea,Herdeiro:2015tia,Chodosh:2015oma} or Proca hair~\cite{Herdeiro:2016tmi}. The former are known as Kerr BHs with scalar hair (KBHsSH). In this paper we have added electric charge to KBHsSH, both considering an ungauged and a gauged scalar field and analysed some basic properties of the solutions. In both cases, their domain of existence is qualitatively similar to the of the uncharged hairy BHs. In particular it is bounded by three curves, corresponding to the solitonic limit (boson stars), extremal hairy BHs, and bald (KN) BHs. In the gauged case, the solitonic limit corresponds to rotating charged boson stars, which hitherto have not been studied in the literature. 

As an example of a novel physical property of these solutions we have considered the gyromagnetic ratio, $g$. This quantity measures how a magnetic dipole moment is induced by the charge and angular momentum of the BH. It is well known the relativistic (Dirac) value holds for the KN BH, $g=2$~\cite{Carter:1968rr}. We have shown that the gyromagnetic ratio of these hairy charged BHs is always $g\leqslant 2$, with equality attained only in the ``bald" case. Thus, the scalar hair leads to a suppression of the magnetic dipole moment. Preliminary work (not reported herein), analysing the electromagnetic field lines of the BHs, suggests that the scalar hair also suppresses the higher electric multipole moments. We hope to give a detailed account of this behaviour in the near future.

There are other interesting applications for these solutions. As an example we mention testing the no-short hair conjecture. As originally stated~\cite{Nunez:1996xv}, this conjecture suggested that, when hair exists around a spherically symmetric BH, the `hair' should extend
beyond $3r_+/2$, where $r_+$ is the areal radius of the event horizon. This radius coincides with the location of the
circular null geodesic for the Schwarzschild solution, which led to an improved version of the conjecture that the hair must extend beyond the null circular orbit of the spacetime~\cite{Hod:2011aa}. For rotating BHs, an analysis of linearized hair suggested the no-short hair conjecture holds for uncharged BHs~\cite{Hod:2016dkn}, but may be violated for electrically charged ones~\cite{Hod:2014sha,Hod:2015ynd}. The latter possibility can be tested using the fully non-linear solutions of the Einstein-Maxwell-Klein-Gordon field equations reported in this paper. 

Finally, we emphasize that all of the analysis presented herein is classical. At the quantum mechanical level, the introduction of electric charge can lead to new phenomena, such as the pair production of oppositely charged particles ($a.k.a.$ Schwinger pair production~\cite{Schwinger:1951nm}). In the context of BH physics, it has been argued that this phenomenon sets un upper bound on the electric field strength outside the horizon and on the BH's electric charge~\cite{1974Natur.247..530Z}. It will be very interesting to analyse the physical properties of KNBHsGSH and KNBHsUSH in relation with this quantum mechanical bound (see $e.g.$~\cite{Hod:2015hga} for a related discussion in the gauged case and test field approximation).

\section*{Acknowledgements}
C. H. and E. R. acknowledge funding from the FCT-IF programme. H.R. is supported by the grant PD/BD/109532/2015 under the MAP-Fis Ph.D. programme. This  work  was  partially  supported  by  the  H2020-MSCA-RISE-2015  Grant  No. StronGrHEP-690904,  and  by  the  CIDMA  project  UID/MAT/04106/2013.  Computations were performed at the BLAFIS cluster, in Aveiro University.

\bigskip

\bibliography{bibtex_calculation}
\bibliographystyle{h-physrev4}

\end{document}